\documentclass[prd, twocolumn, superscriptaddress,floatfix, nofootinbib, preprintnumbers]{revtex4-2}
\pdfoutput=1
\usepackage[utf8]{inputenc}
\usepackage{hyperref}
\hypersetup{colorlinks=true,citecolor=blue}\usepackage{rotating}
\usepackage{array}
\usepackage{amsmath,amssymb,amstext}
\usepackage[normalem]{ulem}
\usepackage{slashed}
\usepackage{booktabs}
\usepackage[pdftex,table,dvipsnames]{xcolor}
\usepackage{multirow}
\usepackage{url}
\usepackage{color}
\usepackage{xspace}
\usepackage{comment}
\usepackage{enumitem}
\usepackage[caption=false]{subfig}
\usepackage{siunitx}
\usepackage{graphicx}
\usepackage{placeins}
\usepackage{layouts}
\usepackage{tikz}

\usepackage{multirow}
\begin{document}

%\preprint{}
\title{Look everywhere effects in anomaly detection}

\author{Marie Hein}
\email{marie.hein@rwth-aachen.de}
\affiliation{Institute for Theoretical Particle Physics and Cosmology, RWTH Aachen University, D-52056 Aachen, Germany}
\author{Benjamin Nachman}
\email{nachman@stanford.edu}
\affiliation{Department of Particle Physics and Astrophysics, Stanford University, Stanford, CA 94305, USA}
\affiliation{Fundamental Physics Directorate, SLAC National Accelerator Laboratory, Menlo Park, CA 94025, USA}
\author{David Shih}
\email{shih@physics.rutgers.edu}
\affiliation{NHETC, Dept.\ of Physics and Astronomy, Rutgers University, Piscataway, NJ 08854, USA}

\begin{abstract}
Machine learning-based anomaly detection methods are able to search high-dimensional spaces for hints of new physics with much less theory bias than traditional searches. However, by searching in many directions all at once, the statistical power of these search strategies is diluted by a variant of the look elsewhere effect.  We examine this challenge in detail, focusing on weakly supervised methods. We find that training and testing on the same data results in badly miscalibrated $p$-values due to the anomaly detector searching everywhere in the data and overfitting on statistical fluctuations. However, if these $p$-values can be calibrated, they may offer the best sensitivity to anomalies, since this approach uses all of the data. 
Conversely, training on half of the data and testing on the other half results in perfectly calibrated $p$-values, but at the cost of reduced sensitivity to anomalies.
Similarly, regularization methods such as early stopping can help with $p$-value calibration but also possibly at the expense of sensitivity. Finally, we find that $k$-folding strikes an effective balance between calibration and sensitivity. Our findings are supported by numerical studies with Gaussian random variables as well as from collider physics using the LHC Olympics benchmark anomaly detection dataset.
\end{abstract}

\maketitle

\section{Introduction}
\label{sec:Introduction}

New physics searches have explored a plethora of specific models of physics beyond the Standard Model (BSM). Recently, model-agnostic anomaly detection (AD) powered by modern machine learning (ML) has complemented traditional analyses, enabling searches for many scenarios simultaneously~\cite{Kasieczka:2021xcg, Aarrestad:2021oeb,Karagiorgi:2022qnh,Belis:2023mqs}.
Many searches of this type already deployed on experimental data~\cite{ATLAS:2020iwa,CMS:2024nsz,ATLAS:2025obc,Gambhir:2025afb}
target overdensity anomaly detection, where data in a region of phase space (the signal region) are compared with a reference (often using a weakly-supervised classifier) and events in overdense subregions are selected.  The number of selected events is then compared to a background prediction and a $p$-value quantifying the statistical compatibility is computed.  The utility of this search strategy is that it can automatically detect anomalies across the full phase space.  At the same time, these strategies must contend with widespread statistical fluctuations. 

In this paper, we explore the hidden effects of the multiple testing problem
 found in overdensity-based AD analyses (henceforth, simply AD analyses).  Such effects are statistical dilutions affecting the sensitivity and/or the accuracy of an AD analysis.  They are a more subtle incarnation of the usual
 \textit{look elsewhere effect} that affects 
classical searches with pre-determined selection criteria. When searches have many, possibly overlapping, signal regions, they 
 suffer from a statistical dilution on the accuracy (mis-calibrated $p$-values) due to a mismatch between local and global significances.  There are well-known approaches to account for it~\cite{Gross:2010qma} and generally involve recalibrating the local $p$-values by a ``trials factor" that very roughly scales with the number of signal regions.
As we will show in this work, a version of the look elsewhere effect impacts AD analyses as well. But the notion of signal regions is more subtle -- the anomaly detector does not search in a predefined and discrete set of signal regions, but in some sense looks everywhere, all at once. Also, we will see that one can trade off sensitivity and accuracy with tweaks to the setup, as will be discussed further below. For these reasons, we dub the issues associated with statistical dilution in the context of AD analyses {\it look everywhere effects}. 

The first AD analyses predated modern machine learning and used a large number of predefined signal regions (see e.g.~\cite{ CMS:2020zjg, ATLAS-CONF-2014-006} and references therein).
They incurred a large trials factor from the look elsewhere effect due to the large number of bins.  It is often said that should a significant excess be identified in one of these bins, then an independent (likely subsequently collected) dataset could be used to focus in on the one bin.  This followup analysis does not have a trials factor.  However, its sensitivity decreases with the number of effectively independent bins from the first search.  In other words, if you add many bins that are known to be not BSM sensitive to the first-round search, then you increase the chance of selecting a statistical fluctuation instead of a bin with actual signal.  This setup highlights three important choices that will reoccur later, including how to pick the anomalous region (here: bin with largest overdensity in the first round), which dataset is used to pick the region (here: the full first round dataset), and which dataset is used for the final analysis (here: the independent dataset).  

Machine learning-based analyses do not rigidly discretize the phase space, but similar features to the binned case are still present.  Like adding useless bins above, adding signal insensitive features to the weakly-supervised classifier training generally reduces the sensitivity \cite{Finke:2023ltw}.  This can be partially recovered by picking ML approaches that are generically more robust to noise~\cite{Finke:2023ltw, Freytsis:2023cjr} or that use pretraining to restrict the function space in some way~\cite{Li:2024htp, Mikuni:2025tar, Bhimji:2025isp, Cheng:2024yig, Cheng:2025ewj}. Classical look elsewhere effects also impact AD analyses, such as when combining multiple approaches \cite{Grosso:2024wjt, Golling:2023yjq, CMS:2024nsz, ATLAS:2025obc} or when explicitly scanning over phase space as in bump hunts.

While recent experimental AD analyses \cite{ATLAS:2020iwa, CMS:2024nsz, ATLAS:2025obc} have performed $p$-value calibration studies of their particular setups, no systematic studies of the underlying statistical trends have been conducted thus far. In this paper, we aim to provide such a systematic study by taking the number of bins (discrete case) or dimensionality (ML case) as given. 
Our focus is on the three questions from earlier.  It is always possible to eliminate the trials factor by dividing the dataset in half and training on one half and testing on the other (as above).  However, a classifier trained with less data will be less sensitive to BSM.  This is the tradeoff mentioned earlier between accuracy and sensitivity.  Hybrid approaches have also been proposed, where the data are split into $k$ equal parts and $k-1$ are used to pick the anomalous region of the $k^\text{th}$ part~\cite{Collins:2018epr,Collins:2019jip}. 
Repeated over all parts, this is called $k$-folding and has been used by the recent experimental AD searches  \cite{ATLAS:2020iwa, CMS:2024nsz, ATLAS:2025obc}.  In practice, it can be difficult to calibrate $p$-values in situ because of the challenge to build validation regions.  Even when calibration is possible, large calibration factors can amplify uncertainties.  Thus, it is often useful to develop approaches that are (nearly) calibrated.

After introducing our statistical procedures in more detail in Sec.~\ref{sec:statistical procedures} and the setup including the classifiers and data sets in Sec.~\ref{sec:setup}, our investigation begins in Sec.~\ref{sec:binnedresults}, with a simple binned version of overdensity detection in the context of a two-dimensional Gaussian toy model. Using 2d histograms of the data and the background template, a single bin is selected based on the highest overdensity score, emulating the scenario from above. The (local) $p$-value is calculated for this bin and reported as the global $p$-value. Although we only calculate this one $p$-value, we still looked in many places (all the bins of the 2d histogram). In this analysis, it is easy to follow the underlying statistical effects and we therefore use it to gain an intuition for these effects. In particular, we discuss how evaluating $p$-values on the training set, on an independent test set or using $k$-fold cross validation affect the calibration of the obtained $p$-values. We then describe how this intuition can be transferred to ML-based analyses that use neural networks (NNs) or Boosted Decision Trees (BDTs) to estimate the overdensity score.

Next, we make a connection between look everywhere effects and 
overtraining. Both effects fundamentally describe an overreliance on statistical fluctuations of the ``training set" instead of learning the fundamental structure of the data. By applying strategies to limit overtraining, we show that this equivalence holds in Sec.~\ref{sec:look elsewhere and overfitting}. We also discuss how these considerations transfer to BDTs.

Finally, in Sec.~\ref{sec:signal sensitivity}, we discuss how choices made to mitigate look everywhere effects in the model setup -- concerning both data usage and architecture choices such as early stopping -- can impact an analysis' signal sensitivity. We show how a nuanced consideration of the different effects may lead to accepting the need for some calibration of the $p$-values in order to achieve an optimal sensitivity. In particular, we find that $k$-fold cross validation can strike a good balance between limiting miscalibration while keeping the sensitivity of an analysis high.

Technical aspects of the $p$-value calculation and fits used to calibrate $p$-values are discussed in Apps.~\ref{app: p values} and \ref{app:LHCO calibration}, respectively. App.~\ref{app:vary k} discusses how varying $k$ in $k$-fold cross validation changes $p$-value (mis-)calibration.

Throughout this work, we assume that the background distribution is known, both for choosing the event selection and for determining $p$-values.  In practice; the background has to be estimated from a combination of data-driven and simulation-based approaches.  Biases in the background can also reduce the analysis sensitivity and/or bias the results.  Our numerical examples do not oversample the background~\cite{Hallin:2021wme}, so there could be some effects from statistical fluctuations in the reference dataset.  Other sources reducing the sensitivity or accuracy are left to future investigations.

\section{Statistical Procedures}\label{sec:statistical procedures}
\label{sec:statproc}

We build our statistical procedures on the principle of Classification Without Labels  (CWoLa)~\cite{Metodiev:2017vrx}, assuming a background-only background template (BT) and a signal region (SR) data set, with probability distributions
\begin{equation}
\begin{split}
    p_\text{BT}(x)&=p_B(x) \\ 
    p_\text{SR}(x)&=f_S\cdot p_S(x)+(1-f_S)\cdot p_B(x), 
\end{split}  
\end{equation}
where $p_{B/S}$ are the background and signal distributions, respectively. The signal fraction in the SR is $f_S$ and $x$ are the features that will be used to find overdensities.  One can therefore obtain a background expectation based on the BT, to which the SR data is then compared.

One approach\footnote{Note that some papers show how to avoid making a cut, including Refs.~\cite{Gambhir:2025afb,Cheng:2025ewj,Kamenik:2022qxs,DAgnolo:2018cun,DAgnolo:2019vbw}.}
in this paradigm of anomaly detection analyses is to define an anomaly score based on the overdensity of the data
\begin{equation}
R(x)={p_\text{SR}(x)\over p_\text{BT}(x)}.
\end{equation}
Then, we put a threshold on $R(x)$ to select a subset of events. We compare this to the background estimation and ask if the distribution of selected events in the data follows the correct distribution (gives calibrated $p$-values under the null hypothesis). 

We will compare a number of ways of estimating $R(x)$ from the data and see to what extent they result in calibrated $p$-values. 
\begin{enumerate}

\item {\it Binning the data:} This works for data in low dimensions (e.g., 2d data in our Gaussian toy example to follow). Here, we estimate $R(x)$ by binning the SR data (i.e.\ density estimation by histogramming) 
and comparing to the expectation of the number of events in the bin coming from the BT (i.e.\ assuming infinite statistics for the BT). The bin with the maximum $R(x)$ is chosen based on the training set. For this bin, a one-sided binomial $p$-value is calculated on an evaluation set (details in App.~\ref{app: p values}), which is reported as the $p$-value of the search.

\item {\it Classifier-based method:}
For the classifier-based procedure, a machine learning classifier is trained on distinguishing BT and SR based on the training set. If this classifier is well trained it approaches the Neyman-Pearson optimal classifier~\cite{Neyman:1933wgr} which is the true likelihood ratio.

To obtain $p$-values based on these classifier scores, we choose a threshold based on the selection efficiency in the BT. Due to statistics limitations of the data sets, we show results for $\epsilon_B=10^{-1}$ and $10^{-2}$ for our Gaussian toy data and add an additional threshold of $10^{-3}$ for the larger LHCO data set. We count how many SR events pass the cut on $R(x)$ and compare it to the expectation from the BT using a negative binomial distribution (see App.~\ref{app: p values}). 

\end{enumerate}

Besides the estimator for $R(x)$, an important consideration is how the data is divided up into training and evaluation samples. We will see that not treating this carefully can result in miscalibration of $p$-values.

\begin{enumerate}[label=(\alph*)]
    \item We use the full statistics of the data set for training and evaluation. This provides the best statistics for both but also has a significant trials factor.
    \item We use 50\% of the data set for training and 50\% for evaluation. This significantly reduces the statistics for the training and evaluation, but avoids a trials factor.
    \item We perform $k$-fold cross validation with $k=5$ in order to use the full statistics for evaluation and limit the statistics reduction in the training set. We train on $k-1$ folds and evaluate on the last fold and repeat this until all $k$ folds have been used as an evaluation set exactly once. The results for all folds are then combined. While there are different strategies of doing this, i.e., combining $p$-values or counts, all of these assume the $k$ results to be independent. We use the common strategy of combining the counts across all folds and calculating a common $p$-value, which was also used in previous weakly supervised searches \cite{ATLAS:2020iwa, CMS:2024nsz, ATLAS:2025obc}.
\end{enumerate}

\section{Setup}
\label{sec:setup}

\subsection{Classifiers}\label{sec:classifiers}
\subsubsection{Neural Network}\label{sec:NN}
For the NN classifier, we use a simple three layer multi-layer perceptron (MLP) implemented in \texttt{Tensorflow} \cite{tensorflow2015-whitepaper} with 64 nodes each and ReLU activation trained using an \texttt{Adam} optimizer \cite{Kingma:2014ad} with a learning rate of $10^{-3}$. We train this classifier with a batch size of 128 for a maximum of 50 epochs either without early stopping and a validation set or with early stopping and a 50-50 training-validation split. The early stopping is performed on the validation loss with a patience of ten epochs and reverts back to the best epoch after the training is stopped. 

\subsubsection{Boosted Decision Tree}\label{sec:BDT}
For the BDT classifier, we use the \texttt{HistGradientBoostingClassifier} from \texttt{scikit-learn} \cite{Pedregosa:2011sk}, a gradient boosted decision tree classifier, which histograms the classification features before training. We use default hyperparameters including early stopping with a patience of ten epochs, where we revert back to the best model. We ensemble ten BDTs with randomized 50-50 training-validation splits. 

\subsection{Data sets}\label{sec:datasets}

\subsubsection{Toy data}\label{sec:toy data}
We use a 2d Gaussian with mean $\mu=0$ and standard deviation $\sigma=1$ as an easily sampled from toy distribution. 

For both binned and ML methods, we sampled 25,000 events in the SR to serve as the data. For case (a) in Sec.~\ref{sec:statproc} (training and evaluation on the same set), the methods were applied directly to this data. For case (b) (training and evaluation on independent sets), we reuse the trainings on these 25,000 SR events to save on compute and apply them to another 25,000 independently sampled events.\footnote{Technically, this doubles the dataset size relative to case (a). However, as we  are not studying signal sensitivity here, and anyway expect calibrated $p$-values in this case, a simplified test was performed to reduce unnecessary computation time.} Finally, for case (c) ($k$-fold cross validation), these 25,000 SR events are split into the $k=5$ folds.

Regarding the BT, for the binned procedure, as stated in Sec.~\ref{sec:statistical procedures}, we use infinite statistics, making use of the fact that we can calculate precise expectation values for this toy distribution. For training the ML classifiers, we use the same statistics for the BT as for the SR data, i.e.\ no oversampling is assumed.

\subsubsection{LHCO data set}\label{sec:LHCO data set}
We use a data set based on the R\&D set \cite{LHCOdataset} from the LHC Olympics 2020 \cite{Kasieczka:2021xcg} as both a more realistic test of the statistical procedures and as a test of the signal sensitivity of the different classifier-based setups and data usage procedures. This data set contains a QCD dijet background and $W'$ resonance signal at $m_{W'}=3.5\,\text{TeV}$, which decays into $X(\rightarrow qq)$ and Y$(\rightarrow qq)$ with masses $m_X=100\,\text{GeV}$ and $m_Y=500\,\text{GeV}$, respectively. 

All events are simulated using \texttt{Pythia 8}~\cite{Sjostrand:2006za,Sjostrand:2007gs} and \texttt{Delphes 3.4.1}~\cite{deFavereau:2013fsa} and clustered using the anti-$k_T$ algorithm~\cite{Cacciari:2005hq} with a jet radius of $R=1$ in Fastjet~\cite{Cacciari:2011ma}. A leading-jet requirement of $p_{\mathrm{T}}>\SI{1.2}{\tera\electronvolt}$ is applied to the data in order to be full efficient with respect to the generator thresholds.  The two highest $p_T$ jets are selected. We use the dijet mass $m_{JJ}$ to obtain only the SR data around the $W'$ resonance between $3.3\,\text{TeV}$ and $3.7\,\text{TeV}$. In practice, one would scan over mass windows and this incurs a classical trials factor.  For the classification, we use a standard feature set also used in Ref.~\cite{Hallin:2021wme}, namely the mass of the lighter jet $m_{J_1}$, the mass differences of the two jets $\Delta m_J=m_{J_2}-m_{J_1}$ and both jets' 21-subjettiness ratios \cite{Thaler:2010tr, Thaler:2011gf}.  

In order to be able to randomly sample iterations of the background-only data set for the calibration tests, we use a set of 10M background events, of which we retain only the approximately 1.2M SR events. We also use the SR signal events, which are approximately 75.3\% of the 100k signal events in the LHCO R\&D set~\cite{LHCOdataset} as well as SR background events from Ref.~\cite{extraLHCOdataset}, which was produced for Ref.~\cite{Hallin:2021wme}.

For the calibration tests, we randomly sample sets of 120k events each for the BT and SR, giving us a similar data set size as given by the standard R\&D set SR. Using this bootstrapping procedure does not give us a full statistical description of the dataset. However, we show that the effect of this is minimal with the size of the bootstrapping dataset we use. The BT and SR events are then used as described in Sec.~\ref{sec:statistical procedures}. 

For the test with signal, we again sample 120k events for the BT. For the SR set, we sample $N_\text{sig,SR}= f_\text{SR}\cdot N_\text{sig}$ signal events, where $f_\text{SR}$ is the fraction of SR events and $N_\text{sig}$ is the number of SR events one would inject into the whole data set\footnote{This definition is made in order for results to be more analogous to other weakly supervised anomaly detection papers using the LHCO R\&D data, where the signal injection is commonly defined on the full data set.}, and sample background events such that the full SR set contains 120k events. These sets are again used for training and validation as described in Sec.~\ref{sec:statistical procedures}. For each signal injection, we train and evaluate ten times with data shuffled before splitting into train and test set or folds with different random selections of the signal events. In order to evaluate how the different data usages affect the training of the classifiers, we also evaluate the signal sensitivity of the different methods on a test set by evaluating the maximum Significance Improvement Characteristic (SIC) value $\epsilon_S/\sqrt{\epsilon_B}$ with a statistics cutoff representing a maximum relative error on the background of 20\%. This test set contain 20k signal events from the R\&D data set and 340k events of the additional SR background events from Ref.~\cite{extraLHCOdataset}. For the $k$-folding, we must make a choice of how to combine multiple classifiers. We choose to base the SIC calculation on the concatenation of the test set classifier scores for each of the $k$ folds.

\section{Results} \label{sec:results}

\subsection{2d Gaussian data}\label{sec:binnedresults}

\begin{figure*}[ht]
        \centering
        \includegraphics[width=\textwidth]{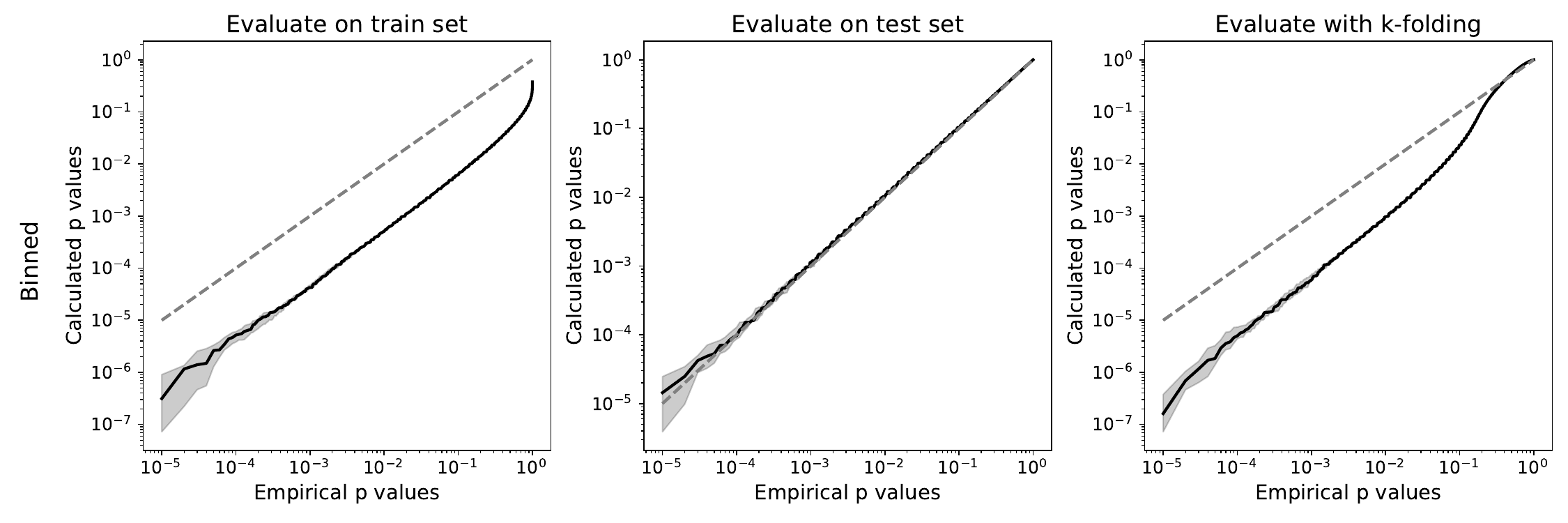}
        \includegraphics[width=\textwidth]{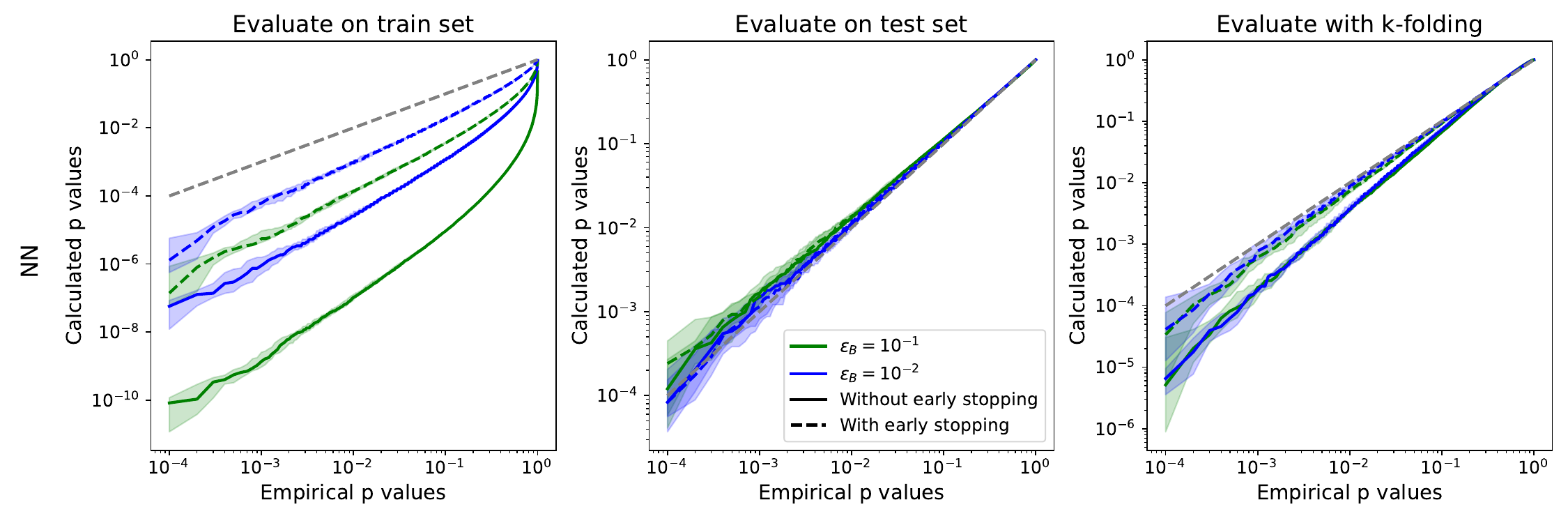}
        \includegraphics[width=\textwidth]{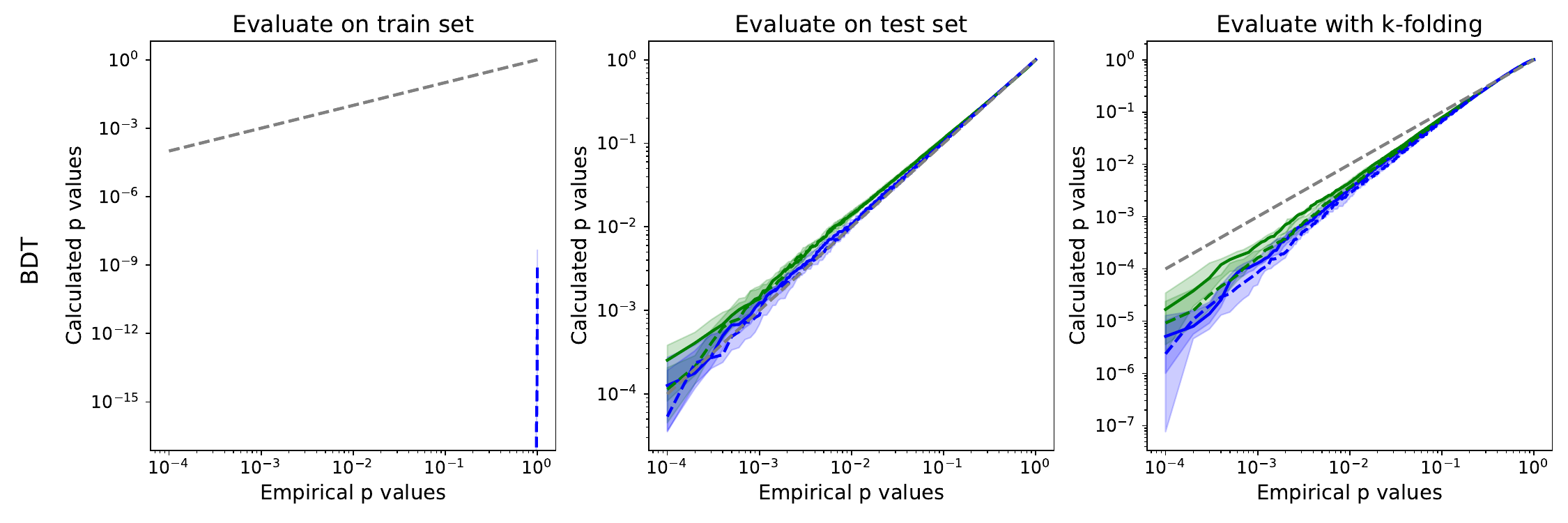}
        \caption{Calibration curves showing empirical cumulative probability and calculated $p$-values observed in the binned classifier (top), the NN with and without early stopping (middle) and BDT with and without early stopping (bottom) for an evaluation on the training set (left), on the test set (middle) and using $k$-folding (right) using the 2d Gaussian toy data.}   
        \label{fig:toydata}
\end{figure*}

\subsubsection{Comparing analysis strategies}

We start with a brief discussion of the simple binned anomaly detection method on the 2d toy Gaussian data, to gain intuition. We divide the data into five bins per dimension between $-2$ and 2, which results in a total of 25 bins. We then pick the bin with the highest overdensity score on the training set. 
 The results are shown in the top row of Fig.~\ref{fig:toydata}. When 
``training", i.e., picking the bin, and evaluating on the same data set as is done in the left panel, we obtain 
miscalibrated $p$-values. This can be interpreted as a look-elsewhere effect: in selecting the most discrepant bin among all the data, we are essentially reporting the minimum local $p$-value as a global $p$-value. 

If we instead evaluate the same bin on an independent test set as is done in the second panel, we obtain calibrated $p$-values as we are no longer subject to the fluctuation of the training set in our evaluation. However, in order to evaluate on an independent data set, assuming a fixed total data set size, we must reduce both the training and evaluation statistics. The effect of this is discussed in the analysis with signal in Sec.~\ref{sec:signal sensitivity}.

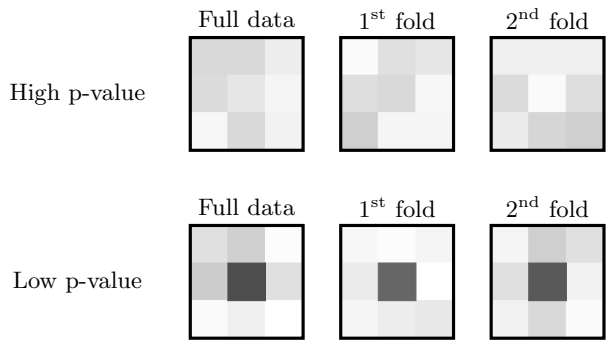
\begin{figure}
    \centering
    \begin{tikzpicture}
    \pgfmathsetseed{6}
    \foreach \n in {0,1,2}{
            \foreach \k in {0,1,2}{
                \foreach \j in {0,1,2}{
                    \pgfmathrandominteger{\r}{0}{20} % random int 0–100
                    \pgfmathsetmacro{\g}{\r/100}
                    \filldraw[color=black!\r] (\k*0.5+\n*2.0,0-\j*0.5) rectangle (0.5+\k*0.5+\n*2.0, -0.5-\j*0.5);
                }
            }
        \draw[very thick] (0+\n*2.0,0) rectangle (1.5+\n*2.0,-1.5);
    }
    \node at (-1.5,-0.75) {High $p$-value};
    \node at (0.75,0.25) {Full data};
    \node at (2.75,0.25) {$1^\text{st}$ fold};
    \node at (4.75,0.25) {$2^\text{nd}$ fold};

        \foreach \n in {0,1,2}{
            \foreach \k in {0,1,2}{
                \foreach \j in {0,1,2}{
                    \pgfmathrandominteger{\r}{0}{20} % random int 0–100
                    \pgfmathsetmacro{\g}{\r/100}
                    \filldraw[color=black!\r] (\k*0.5+\n*2.0,0-\j*0.5-2.5) rectangle (0.5+\k*0.5+\n*2.0, -0.5-\j*0.5-2.5);
                }
            }
        \draw[very thick] (0+\n*2.0,-2.5) rectangle (1.5+\n*2.0,-1.5-2.5);
    }
    \filldraw[color=black!70] (0.5,-3) rectangle (1., -3.5);
    \filldraw[color=black!60] (2.5,-3) rectangle (3., -3.5);
    \filldraw[color=black!65] (4.5,-3) rectangle (5., -3.5);
    
    \node at (-1.5,-3.25) {Low $p$-value};
    \node at (0.75,-2.25) {Full data};
    \node at (2.75,-2.25) {$1^\text{st}$ fold};
    \node at (4.75,-2.25) {$2^\text{nd}$ fold};

    \end{tikzpicture}
    \caption{Illustration of the miscalibration seen in $k$-fold cross validation. The grey-scale represents the excess in each bin.}
    \label{fig:k-fold explanation}
\end{figure}

Finally, the $k$-fold cross validation strategy is shown in the top right panel of Fig.~\ref{fig:toydata}. Here we see that, perhaps surprisingly, despite the $k$-folding, the $p$-values are still miscalibrated. This is because the $p$-values of the individual $k$-folds are actually not independent, due to the following subtlety, which is illustrated for two folds in Fig.~\ref{fig:k-fold explanation}: In the case of relatively small excesses (top row), the largest excess is likely to be diluted when splitting the data into multiple folds. Different bins are then chosen and the $k$-folding serves to mitigate the miscalibration, which we see as a reduced miscalibration for high $p$-values. However, when excesses become large, as is the case in the bottom row of Fig.~\ref{fig:k-fold explanation}, this large excess will likely still be present in multiple folds after splitting the data set. The bin with this large excess is then likely to be selected and evaluated for multiple folds, in extreme cases even for all folds, in which case the miscalibration is not improved compared with the evaluation on the training set, as can be seen in the tails of Fig.~\ref{fig:toydata}. For $k>2$, there is an additional dependence between folds caused by an increasing overlap between training sets. In App.~\ref{app:vary k}, we discuss how varying $k$ impacts the degree of miscalibration.

Having understood the general statistical considerations, we next turn to the case of $p$-value calibration of anomaly detection with ML-based classifiers. The $p$-values corresponding to an NN classifier are shown as solid lines in the middle row of Fig.~\ref{fig:toydata} (the dashed lines are discussed below in Sec.~\ref{sec:look elsewhere and overfitting}). We see that, essentially, the same trends are visible in all three panels as discussed above for the binned analysis. Fundamentally, the statistical treatment of the data in both scenarios does not differ much. Both the binned analysis and the NN classifier select data based on its position in the phase space after evaluating the phase space region with the largest difference between BT and SR, and both calculate the $p$-value by comparing the selection on the BT to the selection on the SR. 
Essentially any configuration of NN parameters taken together with a threshold forms a separate NN ``bin" and $p$-values evaluated for such a fixed parameter configuration and randomly sampled data would be calibrated local $p$-values, just as for a fixed bin in the binned analysis. How the optimal bin or parameter configuration is chosen -- and the look everywhere effect is therefore introduced -- then differs between the binned analysis and the NN, but the effect remains fundamentally equivalent. 

We also see that tighter cuts slightly reduce the degree of miscalibration when evaluating on the training set. This 
makes sense since the miscalibration is due to a bias in the number of signal region events compared to the expectation from the BT, and with tighter cuts we increase the statistical uncertainty, which can cover this bias.

\subsubsection{Connection between the look-everywhere effect and overfitting}
\label{sec:look elsewhere and overfitting}

While we have seen that neural networks experience a look everywhere effect through their training procedure, we now aim to draw a parallel between the look everywhere effect and overfitting. When optimizing an analysis, the ideal bin selection or weight configuration in a NN would rely exclusively on the distributions underlying the data. However, due to finite statistics, any data-driven optimization, be that a data-driven NN training or the selection of the best bin, is also sensitive to the statistical fluctuations of the training set, i.e., the set that the optimization is performed on. It may learn these statistical fluctuations instead of the underlying distributions, which in machine learning-terminology is referred to as overtraining. When this optimization is then evaluated on the same data set, a miscalibration of $p$-values occurs, as was seen in the previous section, and this miscalibration is referred to as the look everywhere effect. 

To support this argument, we use early stopping, a standard technique used to prevent overfitting, to show that it also reduces the strength of the look everywhere effect. This can be seen in the dashed lines in the middle row of Fig.~\ref{fig:toydata}, where we apply the NN to toy data as before but train with early stopping. 
This not only reduces the miscalibration of $p$-values when evaluating on the training set (left panel), 
but also completely removes the miscalibration when evaluating with $k$-fold cross validation (right panel) (see also Ref.~\cite{Collins:2019jip}).

In addition to NN classifiers, we also test BDTs, as they may be more effective than NNs with tabular data. This is especially relevant as their training and score calculation is fundamentally different from NNs and the precise statistical behavior may therefore also differ. Indeed, this is what we can observe in the bottom row of Fig.~\ref{fig:toydata}: while the evaluation on the test set still shows calibrated $p$-values, the miscalibration of the $p$-values is more severe when evaluating on the training set and when evaluating with $k$-fold cross validation miscalibration is observed despite the use early stopping in the training. Apparently, the influence of early stopping on the miscalibrations observed when using $k$-fold cross validation is minimal.

The explanation for this behavior can be found in the difference between how NNs and BDTs determine the classification score an event is given: A NN is a function which calculates -- based on the weights, biases and activation functions -- a specific return value from the input features. BDTs on the other hand divide the event into groups with similar features and for each group return an average value -- in the case of a classification the abundance of a class -- based on the training set. 
Therefore, even if a BDT was initialized randomly (i.e., a set of trees with random cuts on randomly selected features was used), overfitting would occur in the score calculation alone and as such cannot be eliminated entirely through early stopping\footnote{While this is likely the dominant contribution to the miscalibration of the BDT observed here, there are numerous other differences between BDTs and NNs, which could also contribute. For example, the histogramming applied in the BDT we use and the 1d cut-based classification strategy limit the number of possible different phase space selections the BDT, possibly leading to a higher likelihood of overlapping selections. Additionally, BDT trainings at each selection step choose the best possible cut (selecting between all cut values and all features a cut could be performed on), which could also lead to a larger fraction of similar selections compared to the NN as there are no random initializations in a BDT training which can slightly randomize the outcome of a training.}. 

\begin{figure*}[ht]
        \centering
        \includegraphics[width=\textwidth]{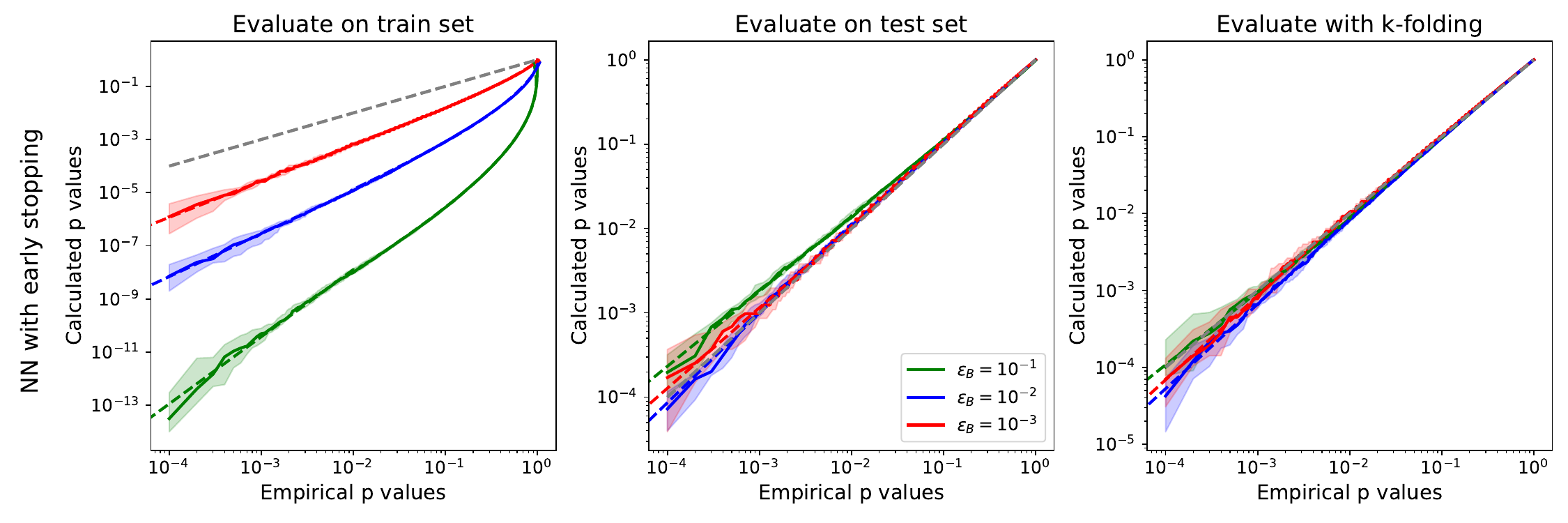}
        \includegraphics[width=\textwidth]{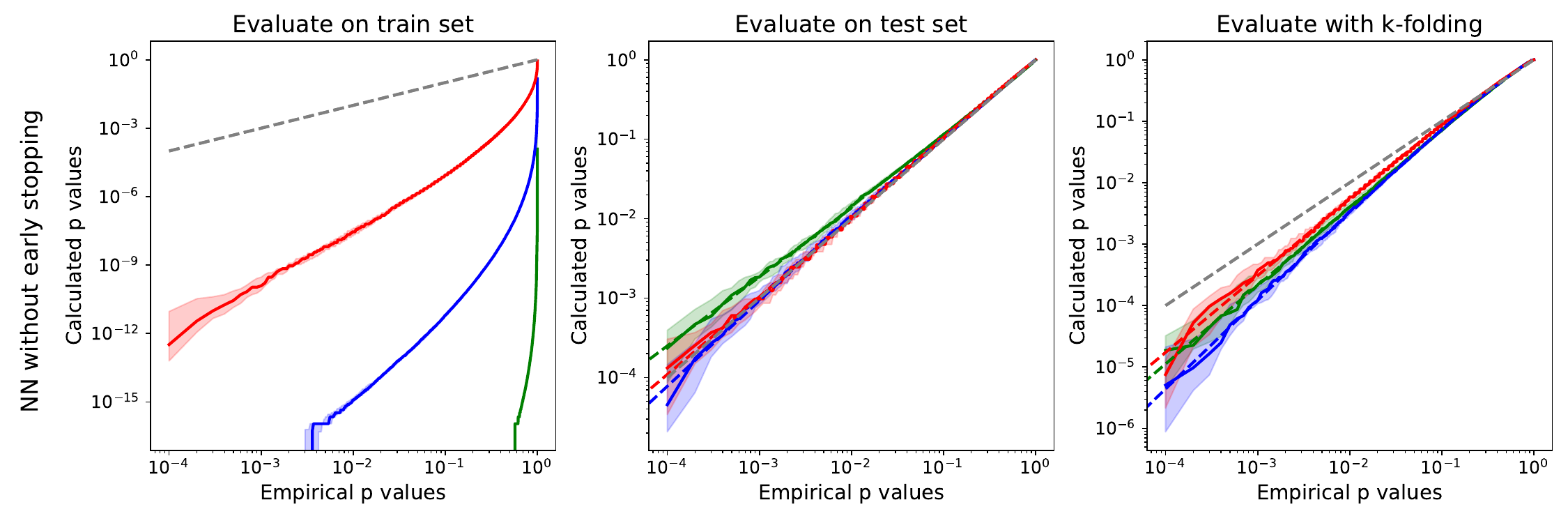}
        \includegraphics[width=\textwidth]{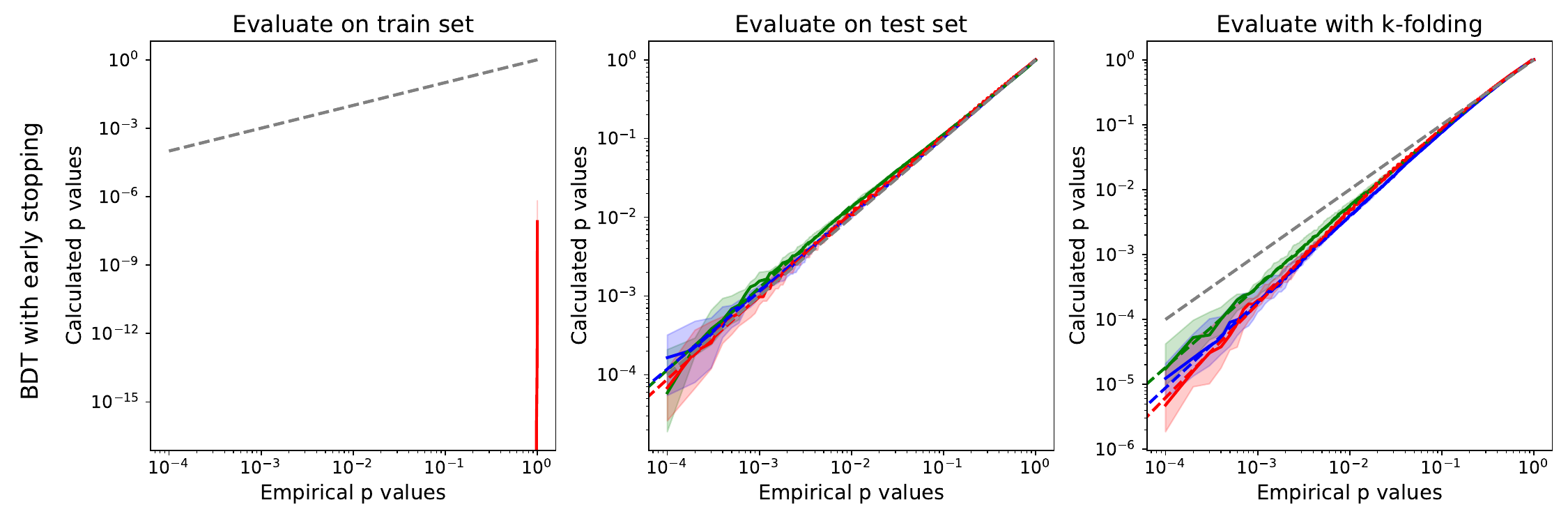}
        \caption{Calibration curves and fits showing empirical cumulative probability and calculated $p$-values observed for the NN with early stopping (top), NN without early stopping (middle) and for the BDT (bottom) for an evaluation on the training set (left), on the test set (middle) and using $k$-folding (right) on LHCO data.}   
        \label{fig:NN BDT LHCO calibration}
\end{figure*}

\begin{figure*}[ht]
        \centering
        \includegraphics[width=\textwidth]{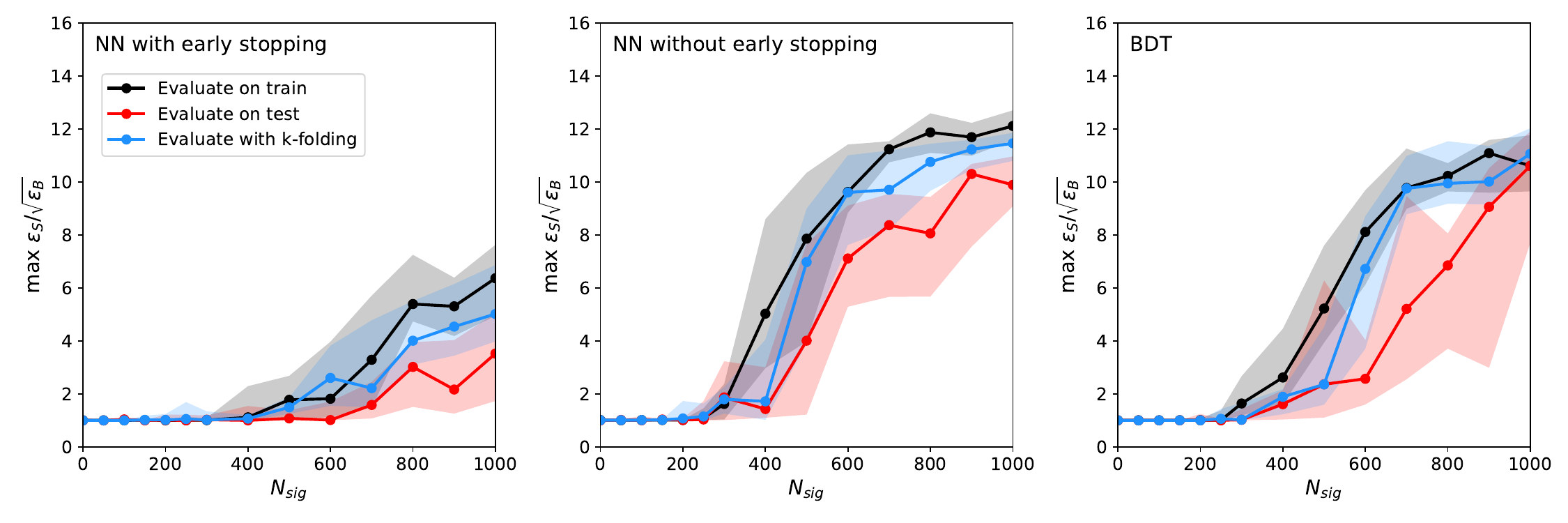}
        \caption{Maximum SIC as a function of the signal injection for classifiers trained on the statistics retained when evaluating on the training set, on the test set and using $k$-folding for NN with early stopping (left), without early stopping (middle) and for the BDT with early stopping (right).}   
        \label{fig:NN BDT LHCO SIC}
\end{figure*}

To say it differently, early stopping for a NN prevents it from overfitting to statistical fluctuations by not allowing the NN parameters to minimize the train loss too much at the expense of the test loss. Meanwhile, early stopping for a BDT means that no additional trees are trained to further refine the phase space selections once the test loss stops improving. This might prevent the BDT ``bins" from becoming too narrowly focused on statistical fluctuations; however, the {\it value} assigned to each BDT bin is still derived from the train set, so it will still be sensitive to statistical fluctuations, despite the early stopping. In this sense the overfitting problem is worse for BDTs than for NNs.

\subsection{LHC Olympics R\&D dataset}
\label{sec:signal sensitivity}

\begin{figure*}[pht]
        \centering
        \includegraphics[width=\textwidth]{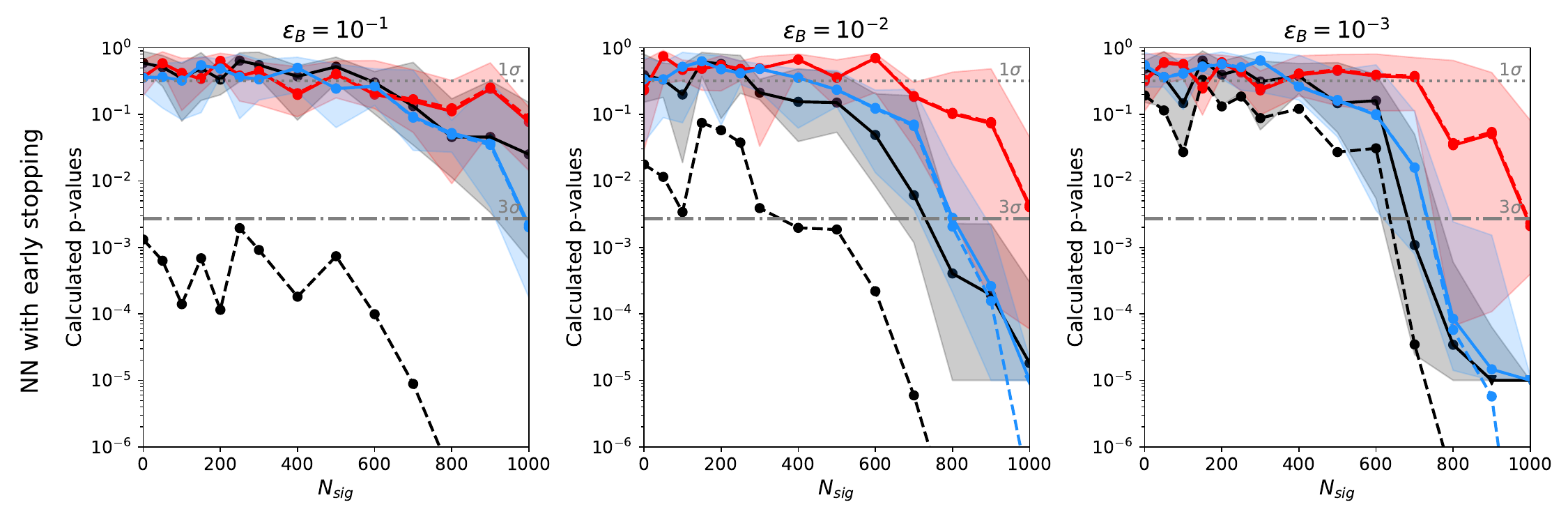}
        \includegraphics[width=\textwidth]{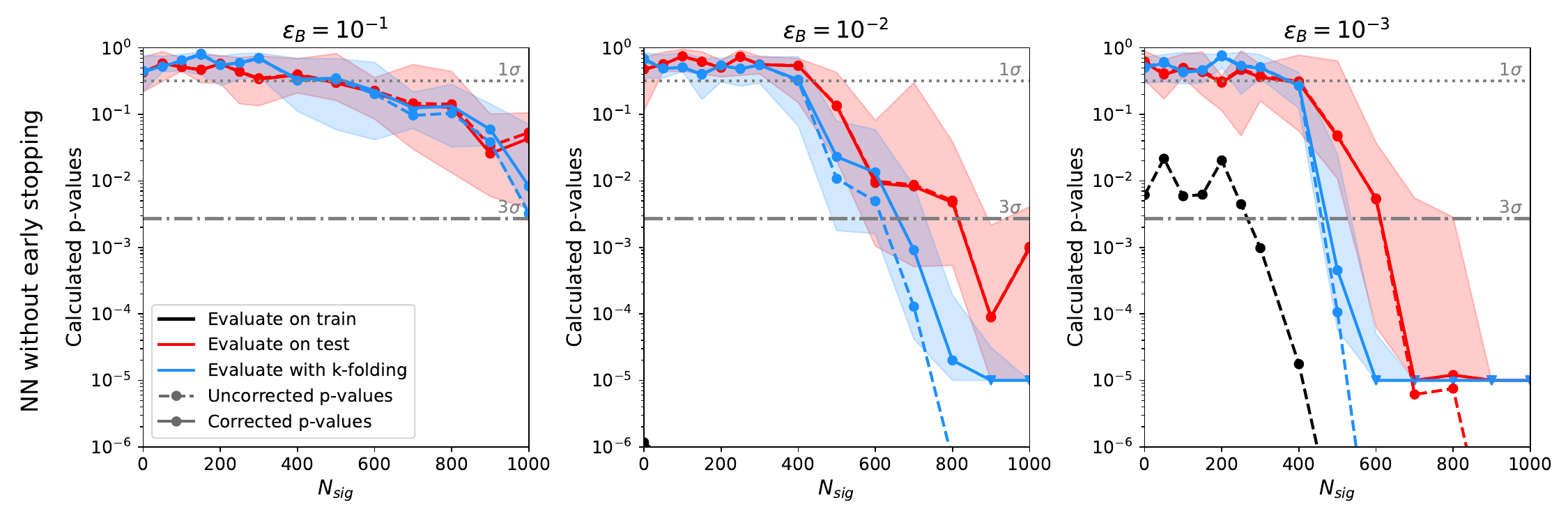}
        \includegraphics[width=\textwidth]{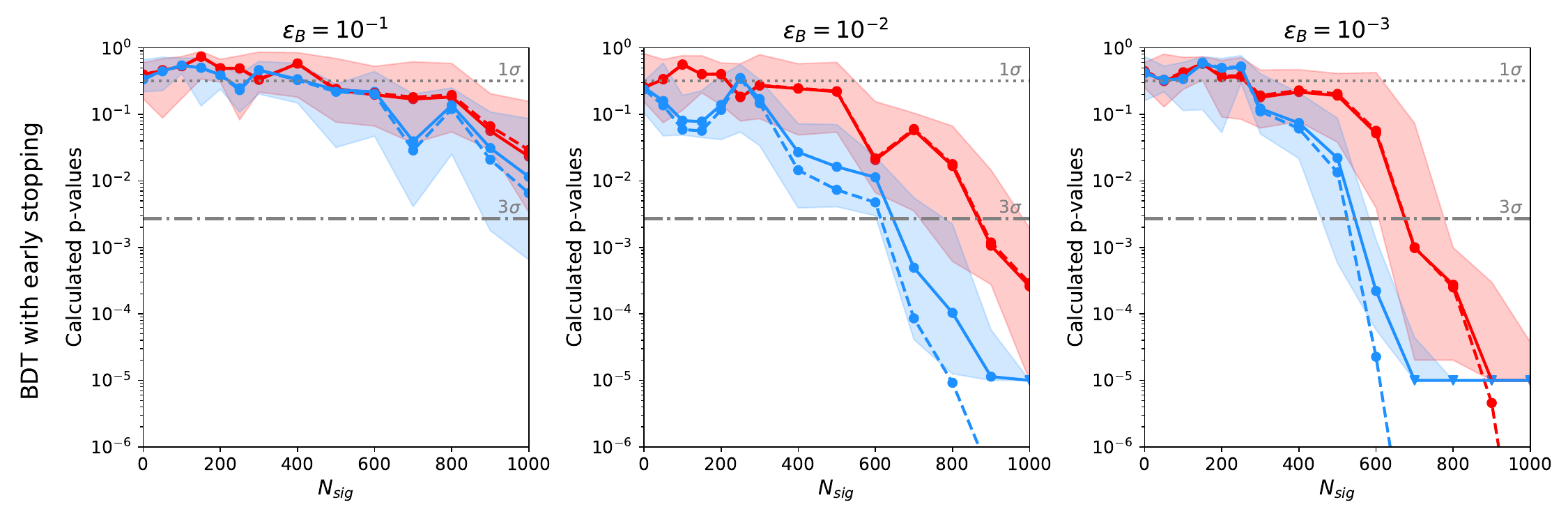}
        \caption{Uncorrected (dashed lines) and corrected (solid lines) $p$-values for different signal injections for NN with early stopping (top), without early stopping (middle) and for the BDT with early stopping (bottom) evaluated on the training set, on the test set and using $k$-folding at three different working points (left, middle, right).}
        \label{fig:LHCO signal pvalues}
\end{figure*}

Next, we turn to a physics case: the LHC Olympics R\&D dataset. First, we repeat the background-only $p$-value calibration tests of the previous subsection. As no significant difference was observed for the BDT between runs with and without early stopping, we continue exclusively with the use of early stopping since this is what has been used in the past (see Ref.~\cite{Finke:2023ltw}). Additionally, as stated in Sec.~\ref{sec:statistical procedures}, we add a lower threshold of $10^{-3}$ as we use a larger data set for these tests. The results are shown in 
Fig.~\ref{fig:NN BDT LHCO calibration}.
We see that the general picture does not change between the toy data example (see Fig.~\ref{fig:toydata}) and a more realistic setup. However, where miscalibration was already observed on the toy data (e.g.\ evaluation on training sets for all), here  
it becomes more severe. This is likely due to the larger feature set allowing for more opportunities for the data to fluctuate, causing larger significances in the selected overfluctuations. 

Second, we turn to the real-life use case of interest -- $p$-value sensitivity and (mis)calibration in the presence of signal.  In order to study the signal sensitivity of an analysis performed using the different options, we need to correct for the observed miscalibration. To do this, we fit calibration curves to the median of $\log_{10} p_{\text{empirical}}(\log_{10} p_{\text{calculated}})$, a detailed description of this fit can be found in App.~\ref{app:LHCO calibration}. In Fig.~\ref{fig:NN BDT LHCO calibration}, these calibration fits are shown as dashed lines. For the NN trained without early stopping and the BDT, fitting the calibration curves of the evaluation on the training set was not possible as $p$-value calculations in some cases returned 0.

Having obtained these calibration fits, we study our different classifiers and data usage strategies with signal. At each signal injection $N_\text{sig}$, we train ten classifiers and evaluate all classifiers on the respective evaluation set (or in the case of the $k$-fold cross validation, perform the entire procedure ten times) as well as evaluate on the test set with background and signal labels as described in Sec.~\ref{sec:datasets} to evaluate the training of the classifier in isolation. The results for the latter can be seen in Fig.~\ref{fig:NN BDT LHCO SIC}, where we plot the maximum of the Significance Improvement Characteristic (SIC), which is defined as $\epsilon_S/\sqrt{\epsilon_B}$ and therefore gives an estimate of the Poisson significance improvement in the Gaussian limit, against the injected number of signal events. To limit statistical fluctuations, we place a statistics cutoff at a 20\,\% statistical error on the background. 

Here, we see the downside of some of the procedures we have used to calibrate $p$-values: In all cases the classifier evaluated on its training set, which does not obtain calibrated $p$-values, performs best as it has the largest training set. The classifier with an independent test set, which always obtains calibrated $p$-values, performs worst as it has the lowest training statistics. Additionally, the NN trained with early stopping performs worst across all settings. The reason for this, as also discussed in Ref.~\cite{Hein:2025uhj} is that the NN often does not pick up on the signal in its first epochs but indeed sometimes significantly later (e.g.\ epoch 40), when overtraining on the background would have already activated the early stopping, which usually reverts the model to one of the first ten epochs. 

The result of an analysis is, however, not solely determined by the classifier's signal sensitivity, but indeed the final $p$-value is a combination of the signal sensitivity and the calibration of $p$-values. These $p$-values we calculate using the scores obtained on the respective evaluation sets and calibrate using the calibration fits described above. This gives us the results shown in Fig.~\ref{fig:LHCO signal pvalues}, where we show both the uncorrected (dashed lines) and corrected $p$-values (solid lines) as a function of the signal injection. It should be noted that where calibration fits were not possible due to the severity of the miscalibration, no corrected curve is shown and that the uncorrected ``evaluation on the training set" curves fall outside the plotting range in some cases. 
Especially interesting here is to consider the miscalibration of the $k$-fold cross validation results, which is in all cases relatively limited. Indeed, it does not result in problematically diverging $p$-values when no signal is injected. Of course, only ten runs are performed here but corrections can also be seen to be minimal at larger signal injections. 

In general we see the same trend emerge here as we already did in the SIC curves of Fig.~\ref{fig:NN BDT LHCO SIC}: While calibrated $p$-values without the application of a trials factor would be ideal, they do not achieve the best sensitivity to new physics signals. While the miscalibration should not get too large, such that a calibration of $p$-values becomes impossible as seen for the BDT and the NN without early stopping when evaluated on the training set, slight miscalibrations caused by $k$-fold cross validation using slightly overfitting models may need to be accepted in order to design a sensitive analysis. However, if miscalibrations, which need to be corrected, are to be avoided entirely, evaluating on a truly independent test set is always a safe option.

\section{Conclusion}
\label{sec:conclusion}
In this paper, we have studied the statistical properties of neural networks and BDTs in the context of anomaly detection. We have shown how anomaly detectors are affected by statistical dilutions that we call look everywhere effects, with connections to classical look elsewhere effects. We have seen how measures taken to reduce the degree of miscalibration in $p$-values result in reduced sensitivity of the networks to new physics signals and can therefore also result in a reduced sensitivity of an analysis. 

While an analysis should always be calibrated, we emphasize the importance of testing this calibration in a setup as close to the final analysis as possible. As we have seen when switching from NNs to BDTs, even seemingly small changes of the analysis setup can have vastly different statistical properties. A careful treatment is vital for the use of machine-learning based analysis setups. In practice, it can be challenging to calibrate with real data and multiple data-driven methods may be required.

Our experiments were performed in a specific weakly supervised anomaly detection scenario using idealized direct background estimation to obtain $p$-values. We expect these results to carry over to both other background estimation methods (e.g.\ interpolating bump hunts) and to other anomaly detection methods, but this deserves further study.
The connection between overfitting and statistical miscalibration may additionally appear beyond anomaly detection and equivalently affect other ML-based scientific analyses such as parameter estimation.

\section*{Acknowledgements}
BN and DS are grateful to Prasanth Shyamsundar for collaboration on an early stage of this project. 
MH thanks Alexander M\"uck for helpful discussions throughout this project. The authors also acknowledge Mark Fanselow for providing the 10M LHCO dijet background events. MH is supported by the Deutsche Forschungsgemeinschaft (DFG, German Research Foundation) under grant 400140256 - GRK 2497: The physics of the heaviest particles at the Large Hadron Collider. BN is supported by the US Department of Energy (DOE) under contract DE-AC02-76SF00515.  DS is supported by DOE grant DOE-SC0010008.
Computations were performed with computing resources granted by RWTH Aachen University under project rwth0934. \\

{\bf Note added:} While this work was being completed, we became aware of the work of \cite{otherADLEE}, which also studies the role of the look elsewhere effect in anomaly searches. 

\section*{Code}
The code for this paper can be found in Ref.~\cite{lookelsewhere-github}.

\appendix

\section{$p$-value calculation}
\label{app: p values}

\subsection{Binned analysis}
In the binned analysis, the following values characterize the result: the total number of signal region events $N_\text{SR}$, the number of signal region event falling into the selected bin $N_{\text{SR,bin}}$ and the expected fraction of background events in the selected bin $p_{\text{exp,bin}}$. In the background-only case, $N_{\text{SR,bin}}$ follows a binomial distribution with $N_\text{SR}$ trials and success probability $p_{\text{exp,bin}}$ such that
\begin{equation}
    N_{\text{SR,bin}} \sim B(N_\text{SR},p_{\text{exp,bin}}).
\end{equation}
Therefore, the $p$-value can be calculated using the cumulative distribution function (CDF) of the binomial distribution $F(N_{\text{SR,bin}};N_\text{SR},p_{\text{exp,bin}})=\text{Pr}(X\le N_\text{SR},p_{\text{exp,bin}})$. We calculate a one-sided $p$-value, taking only excesses into account. Therefore, the $p$-value is given by 
\begin{equation}
    p  = 1 - F(N_{\text{SR,bin}}-1;N_\text{SR},p_{\text{exp,bin}}).
\end{equation}

\begin{figure*}[ht]
        \centering
        \includegraphics[width=\textwidth]{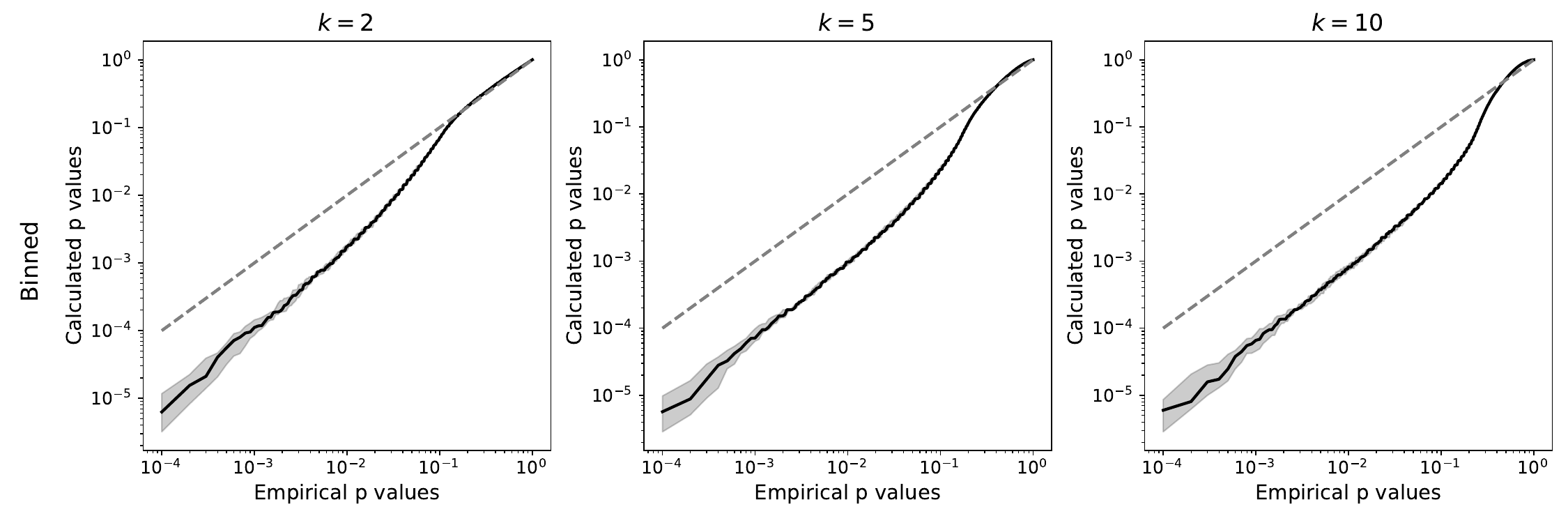}
        \includegraphics[width=\textwidth]{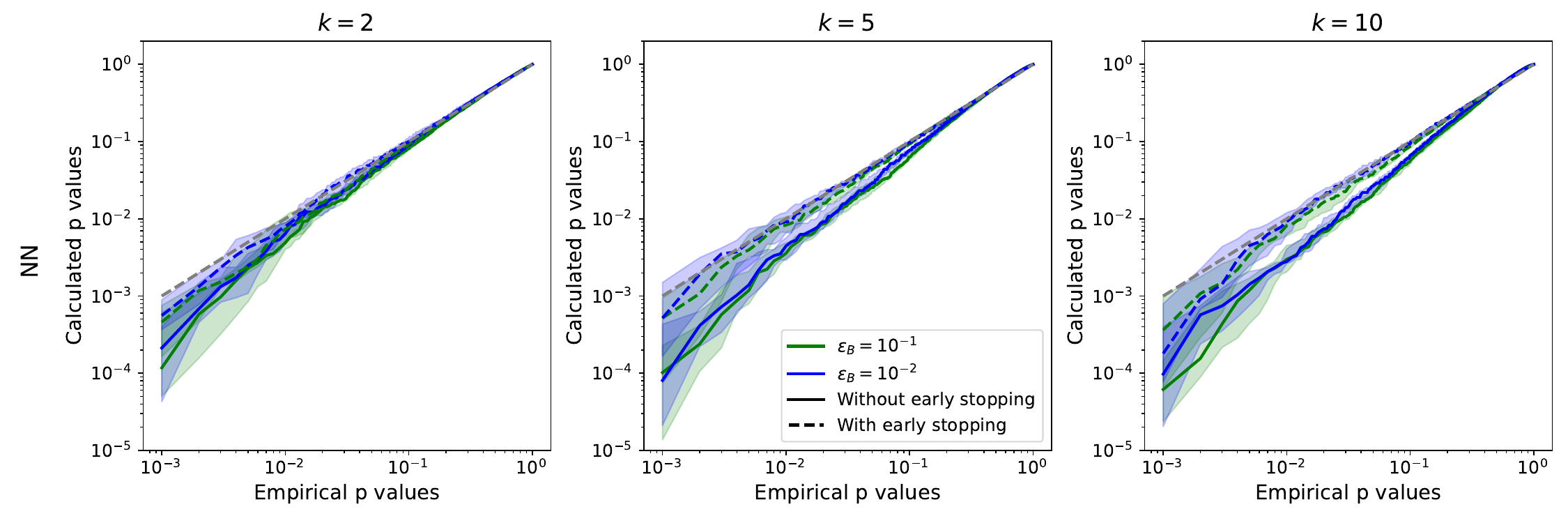}
        \includegraphics[width=\textwidth]{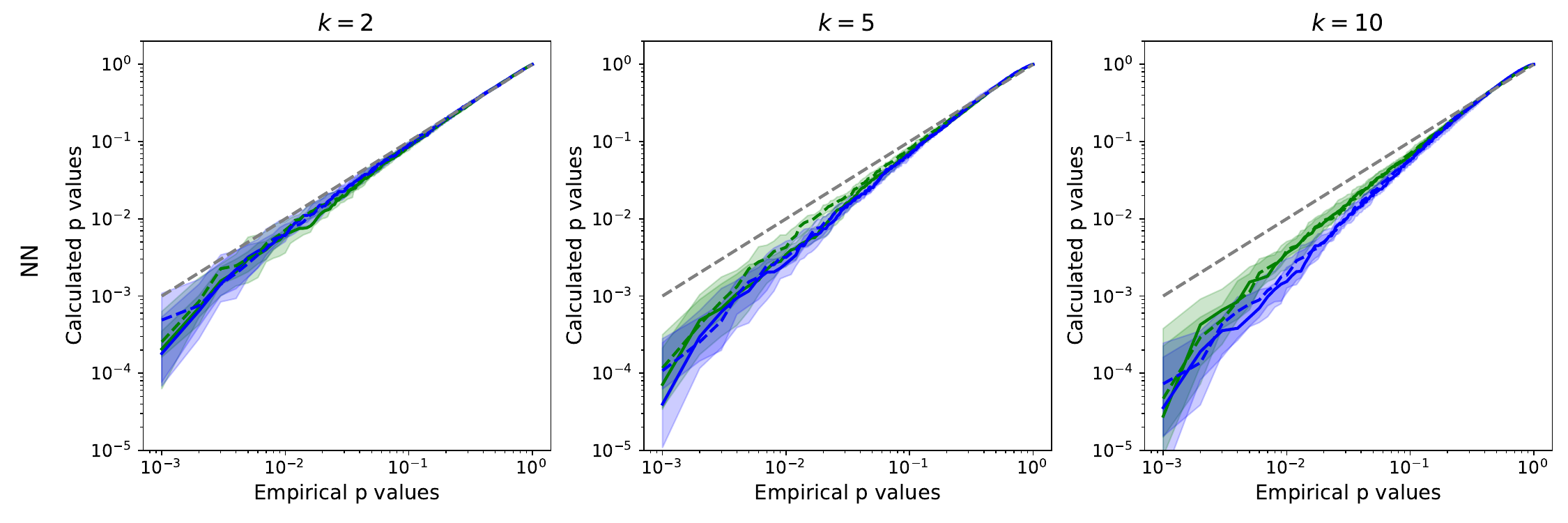}
        \caption{Calibration curves showing empirical cumulative probability and calculated $p$-values observed in the binned classifier (top), the NN with and without early stopping (middle) and BDT with and without early stopping (bottom) for an evaluation using $k$-folding for $k=2$ (left), $k=5$ (middle) and $k=10$ (right using the 2d Gaussian toy data.}   
        \label{fig:toydata vary k}
\end{figure*}

\subsection{Classifier analysis}
For the classifier, the $p$-value depends on the number of signal region events selected by the classifier $N_{\text{SR}}'$ and the number of background template events selected by the classifier $N_\text{BT}'$. Taken in isolation, both $N_\text{BT}'$ and $N_\text{SR}'$ follow a Poisson distribution with the common expectation value $\lambda$. However, $\lambda$ is unknown and can only be estimated using $N_\text{BT}'$. For this, we choose
\begin{equation}
    \lambda \sim \Gamma(N_\text{BT}', 1),
\end{equation}
i.e., that $\lambda$ be drawn from a Gamma distribution, as the Gamma distribution is the conjugate prior of the Poisson distribution. Additionally, we can assume 
\begin{equation}
    N_\text{SR}' \sim \text{Poisson}(\lambda).
\end{equation}
The full distribution is therefore given by the marginalization over $\lambda$ with 
\begin{equation}
    P(N_\text{SR}'|N_\text{BT}') = \int^{\infty} _0 \text{Poisson}(N_\text{SR}'|\lambda)\Gamma(\lambda|N_\text{BT}')d\lambda,
\end{equation}
which is a negative binomial distribution with $p=1/2$ and with $r=N_\text{BT}'$, which are obtained from the precise parametrization of the prior on $\lambda$. Using the negative binomial distribution's CDF, the $p$-value can then be evaluated as was explained for the binned case above. 

\section{Varying $k$ in $k$-fold cross validation}
\label{app:vary k}
In Sec.~\ref{sec:binnedresults}, we discussed the trade-off between a larger statistical dilution and a larger overlap between training sets when increasing $k$ for $k$-fold cross validation. In Fig.~\ref{fig:toydata vary k}, we show how this trade-off manifests in tests of our different classifiers on the 2d Gaussian toy data. In all cases (binned, NN, BDT), we observe the same consistent trend, namely that miscalibration increases as we increase $k$. As such, it seems that the increasing overlap between training sets is more impactful than the increased statistical dilution.

\section{Details on the calibration fits}
\label{app:LHCO calibration}

Here, we explain the calibration fits, which are shown in Fig.~\ref{fig:NN BDT LHCO calibration} and used to calibrate the $p$-values in Fig.~\ref{fig:LHCO signal pvalues}, in detail: We fit the calibration curves to the medians of $\log_{10} p_{\text{empirical}}(\log_{10} p_{\text{calculated}})$ to ensure very low $p$-values are fitted correctly. We fit the spectrum in two parts: First, we fit the  \texttt{UnivariateSpline} from \texttt{scikit-learn}~\cite{Pedregosa:2011sk} to the whole spectrum. These splines tend to diverge in the tails. Therefore, we also fit a linear curve starting at empirical $p$-values of $10^{-2}$, below which we only observe a linear behavior in all tested scenarios. We then switch between the two fits at $10^{-2}$. We limit extrapolation and only allow the fit to return corrected $p$-values down to $\min(p_\text{empirircal})/10$. This fit method works well overall as a stable interpolation between the $p$-values obtained in the calibration tests, as can be seen by eye in Fig.~\ref{fig:NN BDT LHCO calibration}, but was not evaluated rigorously.

\bibliographystyle{apsrev4-1}
\bibliography{HEPML, other}
\end{document}